# Bianchi Type V Matter Filled Universe with Varying Lambda Term in General Relativity

Anil Kumar Yadav

Department of Physics, Anand Engineering College, Keetham, Agra -282 007, India

E- Mail:  abanilyadav@yahoo.co.in

**Abstract:** Bianchi type V matter filled universe with varying $\Lambda$ in general relativity are investigated by using the law of variation for the generalized mean Hubble parameter. This yields the constant value of deceleration parameter and generates two types of solutions for the average scale factor one is of power law type and other is of exponential type. The cosmological constant is found to be a decreasing function of time, which is supported by results from recent type Ia supernovae observations. Also it has been found that cosmological constant $\Lambda$ (t) affects entropy. Some physical and geometric behaviour of the models are discussed.

Key words : Cosmology, Hubble's parameters, Variable cosmological constant and Bianchi type V universe.

PACS :  98.80.Es, 98.80. - k

## 1. Introduction

The study of Bianchi type V cosmological models create more interest as these models contain isotropic special cases and permit arbitrary small anisotropy levels at some instant of cosmic time. As an anisotropic cosmological models, cosmologists generally consider Bianchi type I space-times, which are the simplest generalizations of FRW models. There are still a few other models that describe an anisotropic space-time and generate particular interest among physicists such as Lorenz and Petzold [1], Singh and Agrawal [2], Ibanez et al [3], Berger [4], Marsha [5], Socorro and Medina [6]. Among different models Bianchi type V universes are the natural generalization of the open FRW model, which are eventually become isotropic. A number of authors such as  Franswerth [7], Collins [8], Maartens and Nel [9],Wainwright et al [10], Beesham [11], Maharaj and Beesham [12], Shri Ram [13] and Camci et al [14] have studied bianchi type V models in differents Physical contexts. Christodoulakis et al [15,16] have studied untilled diffuse matter Bianchi V universes with perfect fluid and scalar field coupled to perfect fluid sources obeying a general equation of state.

Following the work of Saha [17], Singh and Chaubey [18, 19] have presented a quadrature form of metric functions for Bianchi type V model with perfect fluid and viscous fluid.

In resent years the solution of Einstein's field equation for homogeneous and anisotropic Bianchi type models have been studied by several authors such as Hajj-Boutros [20, 21], Mazumdar [22] in different text. Solutions of field equations may also be generated by applying the law of variation for Hubble's parameter, which was initially proposed by Berman [23] for FRW models. The main feature of this law is that it yields constant value of deceleration parameter. The cosmological models with constant deceleration parameter have been studied by Maharaj and Naidoo [24], Johri and Desikan [25], Singh and Desikan [26] in different theories of FRW and Bianchi type I models. Recently Singh and Kumar [27] extended Berman works to study anisotropy Bianchi type II space-time models by formulating the law of variation for Hubble's parameters.

Models with a relic cosmological constant $\Lambda$ have received considerable attention recently among researchers for various reasons [28-32]. Some of the resent discussions on the cosmological constant problem and on cosmology with time varying cosmological constant by Ratra and Peebles [33] and Dolgov [34-36] pointed out that in the absence of any interaction with matter or radiation, the cosmological constant remains a "constant". however in the presence of interactions with matter or radiation, a solution of Einstein's equations and the assumed equation of covariant conservation of stress-energy with a time varying $\Lambda$ can be found. For these solutions, conservation of energy requires a decrease in the energy density of the vacuum component to be compensated by a corresponding increase in the energy density of matter or radiation. Earlier researchers on this topic, are contained in Zeldovich [37]. Resent cosmological observations by High-z Supernova Team and Supernova Cosmological project (Garnavich et al [38], Perlmutter et al [39], Rices et al [40], Schmidt et al [41]) strongly favour a significant and positive $\Lambda$ with the magnitude $\Lambda(G\hbar/c^3) \approx 10^{-123}$. Their findings arise from a study of more than 50 type Ia Supernovae with redshifts in the range $0\cdot 10 \leq z \leq 0\cdot 83$ and suggest Friedman models with negative pressure matter such as the cosmological constant, domain walls or cosmic strings. The main conclusion of these observations on magnitude and redshift of type Ia supernovae suggests that the expansions of the universe may be an accelerating one with a large function of cosmological density in the form of the cosmological $\Lambda$-term.

In this paper I extended the work of Singh, Shri Ram and Zeyauddin [42] in the presence of interactions with matter or radiation to specially homogeneous and totally anisotropic Bianchi type V models with perfect fluid as source. In section 2, we present the field equations and in sec.3 we present the law of variation of Hubble's parameter for this space-time that yields the constant value of deceleration parameter. In sec.4, we presented the exact solution of field equations and in sec. 5, we present discussion and concluding remaks.

## 2. Field Equations:

For Bulk Viscous fluid the usual energy momentum tensor is modify by addition of term

$$T_{ij}^{(vac)} = -\Lambda(t) g_{ij}, \qquad (1)$$

Where $\Lambda(t)$ is the cosmological term and $g_{ij}$ is the metric tensor. Thus the new energy momentum tensor is given by

$$T_{ij} = (p + \rho)u_i u_j - p g_{ij} - \Lambda(t) g_{ij}, \qquad (2)$$

Where p and ρ are the energy density and pressure of the cosmic fluid, and $u_i$ is the fluid four velocity such that $u^i u_i = 1$.
We consider the space-time metric of spatially homogeneous Bianchi type v of the form

$$ds^2 = dt^2 - A^2(t)dx^2 - e^{2\alpha x}\left[B^2(t)dy^2 + C^2(t)dz^2\right] \qquad (3)$$

where α is constant. For the energy momentum tensor (2) and Bianchi type V space time (3), Einstien's field equations

$$R_{ij} - \frac{1}{2}R g_{ij} = -8\pi T_{ij} \qquad (4)$$

Yield the following five independent equations

$$\frac{A_{44}}{A} + \frac{B_{44}}{B} + \frac{A_4 B_4}{AB} - \frac{\alpha^2}{A^2} = -8\pi(p + \Lambda), \qquad (5)$$

$$\frac{A_{44}}{A} + \frac{C_{44}}{C} + \frac{A_4 C_4}{A_4 C_4} - \frac{\alpha^2}{A^2} = -8\pi(p + \Lambda), \qquad (6)$$

$$\frac{B_{44}}{B} + \frac{C_{44}}{C} + \frac{B_4 C_4}{BC} - \frac{\alpha^2}{A^2} = -8\pi(p + \Lambda), \qquad (7)$$

$$\frac{A_4 B_4}{AB} + \frac{A_4 C_4}{AC} + \frac{B_4 C_4}{BC} - \frac{3\alpha^2}{A^2} = -8\pi(\rho - \Lambda), \qquad (8)$$

$$2\frac{A_4}{A} - \frac{B_4}{B} - \frac{C_4}{C} = 0, \qquad (9)$$

Here and what follows the suffix 4 by the symbol A, B, C denotes differentiation with respect to t. Taking into account the conservation equation, we have

$$\rho_4 + (\rho + p)\left(\frac{A_4}{A} + \frac{B_4}{B} + \frac{C_4}{C}\right) = 0 \qquad (10)$$

## 3. Model and law of variation for Hubble's parameter

The law of variation for the generalized mean Hubble parameter in the case of a spatially homogeneous and anisotropic for Bianchi type V space time metric that yields a constant value of deceleration parameter. We define the following physical and geometrical parameters to be used in formulating the law and further in solving the Einstein's field equations for the metric (1).

The average scale factor a of the Bianchi type V model is given by

$$a = (ABC)^{1/3} \qquad (11)$$

The spatial volume V is given by

$$V = a^3 = ABC \tag{12}$$

We defined the generalized mean Hubble's parameter H as

$$H = \frac{1}{3}(H_1 + H_2 + H_3) \tag{13}$$

Where $H_1 = \frac{A_4}{A}, H_2 = \frac{B_4}{B}$ and $H_3 = \frac{C_4}{C}$ are the directional Hubble's parameters in the direction of x,y and z respectively. The suffix 4 denotes differentiation with respect to cosmic time t.

From equation (11)-(13), we obtain the following relation

$$H = \frac{1}{3}\frac{V_4}{V} = \frac{a_4}{a} = \frac{1}{3}\left(\frac{A_4}{A} + \frac{B_4}{B} + \frac{C_4}{C}\right) \tag{14}$$

The physical quantity of observational interest in Cosmology are the expansion scalar $\theta$, shear scalar $\sigma^2$ and the average anisotropy parameter $A_m$. All are defined as follows

$$\theta = u^i_{;i} = \left(\frac{A_4}{A} + \frac{B_4}{B} + \frac{C_4}{C}\right) \tag{15}$$

$$\sigma^2 = \frac{1}{2}\sigma_{ij}\sigma^{ij} = \frac{1}{3}\left[\theta^2 - \frac{A_4 B_4}{AB} - \frac{A_4 C_4}{Ac} - \frac{B_4 C_4}{BC}\right], \tag{16}$$

$$A_m = \frac{1}{3}\sum_{i=1}^{3}\left(\frac{\Delta H_i}{H}\right)^2 \tag{17}$$

Where $\Delta H_i = H_i - H$ (i=1,2,3)

Since the line element (1) is completely characterized by Hubble parameter H, therefore let us consider that mean Hubble parameter H is related to the average scale factor a by the relation

$$H = ka^{-n} = k(ABC)^{-n/3} \tag{18}$$

Where k (> 0) and n ($\geq 0$) are constant. Such type of relation have already been considered by Berman[23] for solving FRW models. Later on, many authors have studied flat FRW and Bianchi type models by using the special law of Hubble parameter that yields constant value of deceleration parameter. The deceleration parameter (q) is defined as

$$q = \frac{-a_{44}a}{a_4^2} \tag{19}$$

From equation (14) and (18), we get

$$a_4 = ka^{-n+1} \tag{20}$$

$$a_{44} = -k^2(n-1)a^{-2n+1} \tag{21}$$

From equation (19), (20) and (21), we get

$$q = n - 1 \tag{22}$$

Thus we see that q is constant. The sign of q indicates whether the model inflates or not. The positive sign of q (n >1) corresponds to standard decelerating model whereas the negative sign $-1 \leq q \leq 0\, ie\, (0 \leq n \leq 1)$ indicates inflation. It may be noted that though the current observations of SNe Ia and CMBR favor accelerating models (q < 0), but they do not altogether rule out the decelerating ones which are also consistent with these observations.

From equation (20), we obtain the law of average scale factor a as

$$a = (nkt + c_1)^{1/n}, \quad \text{for } n \neq 0 \tag{23}$$

And

$$a = c_2 \exp(kt), \quad \text{for } n = 0 \tag{24}$$

Where $c_1$ and $c_2$ are the constants of integration. From eq. (23) it is clear that the condition for expansion of universe is n = q + 1 > 0.

## 4. Solution of field equations

Field equations (5) – (8) and (10) can be written in terms of H, $\sigma^2$ and q as

$$8\pi(p + \Lambda) = H^2(2q - 1) - \sigma^2 + \frac{\alpha^2}{A^2}, \tag{25}$$

$$8\pi(\rho - \Lambda) = 3H^2 - \sigma^2 - \frac{3\alpha^2}{A^2}, \tag{26}$$

$$\rho_4 + 3(\rho + p)H = 0, \tag{27}$$

Integrating equation (9) and absorbing the integration constant into B or C, without loss of generality, we obtain

$$A^2 = BC \tag{28}$$

We now present the quadrature form of Einstein's field equations (5) –(9). Subtracting (7) from (6), one finds the following relation between A and B

$$\frac{A}{B} = d_1 \exp\left(k_1 \int \frac{dt}{a^3}\right) \tag{29}$$

Analogically, we find the other relations

$$\frac{B}{C} = d_2 \exp\left(k_2 \int \frac{dt}{a^3}\right), \tag{30}$$

$$\frac{C}{A} = d_3 \exp\left(k_3 \int \frac{dt}{a^3}\right), \tag{31}$$

Where $d_1$, $d_2$, $d_3$, and $k_1$, $k_2$, $k_3$, are constants of integration, obeying

$$d_1 d_3 = d_2^{-1}, \qquad k_1 + k_2 + k_3 = 0 \tag{32}$$

In view of equation (32), we obtain the metric functions from (29)–(31) explicitly as follows [ Saha[17], Singh and Chaubey[19]]

$$A(t) = (l_1)^{1/3} a \exp\left(\frac{K_1}{3} \int \frac{dt}{a^3}\right), \tag{33}$$

$$B(t) = (l_2)^{1/3} a \exp\left(\frac{K_2}{3} \int \frac{dt}{a^3}\right), \tag{34}$$

$$C(t) = (l_3)^{1/3} a \exp\left(\frac{K_3}{3} \int \frac{dt}{a^3}\right), \tag{35}$$

Where $K_1 = k_1 - k_3$, $K_2 = -2k_1 - k_3$, $K_3 = k_1 + 2k_3$,

$$l_1 = \sqrt[3]{\frac{d_1}{d_3}}, \quad l_2 = \sqrt[3]{\frac{1}{d_1^2 d_3}}, \quad l_3 = \sqrt[3]{d_1 d_3^2}.$$

From equation (33) – (35), it is clear that for $a = (nkt + c_1)^{1/n}$ with n > 0, the exponent tends to unity for large value of t and the anisotropic model becomes isotropic.

From equation (28) and (33)-(35), we obtain
$K_1 = 0$, $K_2 = -K_3 = K$, $l_1 = 1$, $l_2 = l_3^{-1} = M^3$,
Where K and M are constants. Now the equation (33)-(35) can be written as

$$A(t) = a, \tag{36}$$

$$B(t) = Ma \exp\left(\frac{K}{3} \int \frac{dt}{a^3}\right), \tag{37}$$

$$C(t) = M^{-1} a \exp\left(-\frac{K}{3} \int \frac{dt}{a^3}\right), \tag{38}$$

Thus the metric functions are represented explicitly in terms of average scale factor a. It is to be noted that once we get the value of a, we can find the metric functions. Many authors have tried to find the solutions the quadrature equation (36)-(38) by using different techniques. Here we solve equation (36)-(38) by using the average scale factor as obtained in equation (23) and (24) for $n \neq 0$ and $n = 0$, respectively by the assumption of equation (18), which have physical interest to describe the decelerating and accelerating universe.

### 4.1 When $n \neq 0$

Using equation (23) into (36)-(38), the solutions for metric functions can be written as

$$A(t) = (nkt + c_1)^{\frac{1}{n}}, \tag{39}$$

$$B(t) = M(nkt + c_1)^{\frac{1}{n}} \times \exp\left[\frac{K}{3k(n-3)}(nlt + c_1)^{(n-3)/n}\right], \tag{40}$$

$$C(t) = M^{-1}(nkt + c_1)^{\frac{1}{n}} \times \exp\left[-\frac{K}{3k(n-3)}(nlt + c_1)^{(n-3)/n}\right], \tag{41}$$

Where $n \neq 3$.

Hence metric (3) reduces to the new form

$$ds^2 = dt^2 - (nkt + c_1)^{\frac{2}{n}} dx^2 - e^{2\alpha x}\left[\begin{array}{l} M^2(nkt+c_1)^{\frac{2}{n}} \exp\left(\frac{K}{3k(n-3)}\left(nkt+c_1\right)^{(n-3)/n}\right) dy^2 + \\ M^{-2}(nkt+c_1)^{\frac{2}{n}} \exp\left(-\frac{K}{3k(n-3)}(nkt+c_1)^{\frac{(n-3)}{n}}\right) dz^2 \end{array}\right], \tag{42}$$

The pressure and density for model (42) is given by

$$8\pi(p + \Lambda) = (2n-3)k^2(nkt+c_1)^{-2} - \frac{K^2}{9}(nkt+c_1)^{-\frac{6}{n}} + \alpha^2(nkt+c_1)^{-\frac{2}{n}} \tag{43}$$

$$8\pi(\rho - \Lambda) = 3k^2(nkt+c_1)^{-2} - \frac{K^2}{9}(nkt+c_1)^{-\frac{6}{n}} - 6\alpha^2(nkt+c_1)^{-\frac{2}{n}} \tag{44}$$

For complete determinacy of the system, we consider a perfect gas equation of state

$$p = \gamma\rho, \qquad 0 \leq \gamma \leq 1. \tag{45}$$

Equation (43), with the use of (45) and (44), reduces to

$$8\pi(1+\gamma)\rho = 2nk^2(nkt+c_1)^{-2} - \frac{2K^2}{9}(nkt+c_1)^{-\frac{6}{n}} - 2\alpha^2(nkt+c_1)^{-\frac{2}{n}} \tag{46}$$

Eliminating $\rho(t)$ between equation (44) and (46), we obtain

$$8\pi(1+\gamma)\Lambda = (2n-1-\gamma)k^2(nkt+c_1)^{-2} - (1-\gamma)\frac{K^2}{9}(nkt+c_1)^{-\frac{6}{n}} + (3+3\gamma-2)\alpha^2(nkt+c_1)^{-\frac{2}{n}} \tag{47}$$

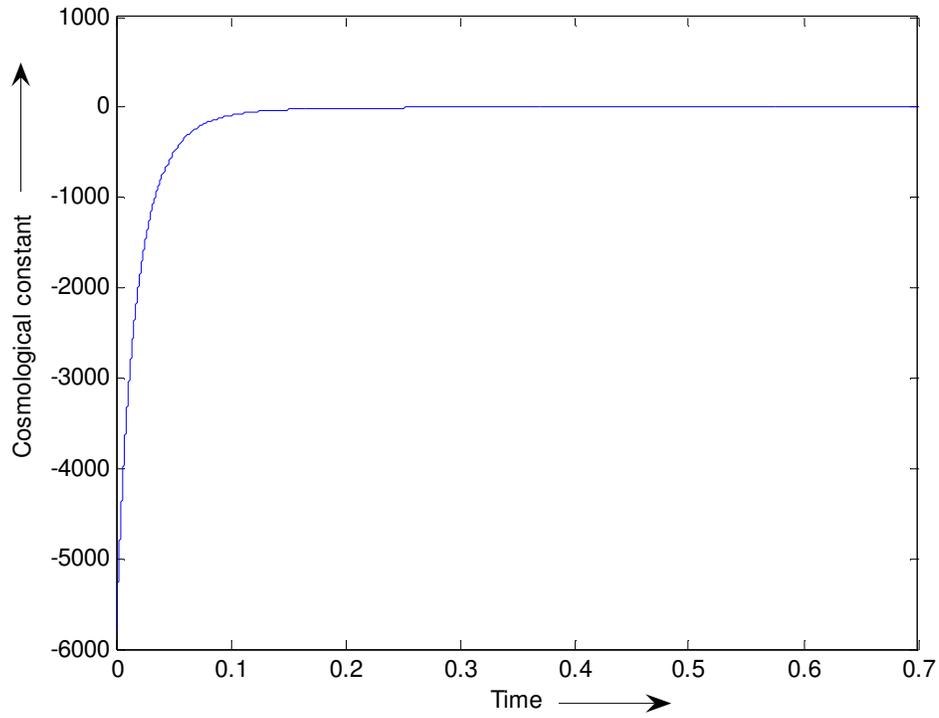

Fig.1. The plot of cosmological constant $\Lambda$ versus time t for the model (42) with parameters $\gamma = 0.5$, $\alpha = 0.2$, $n = 1$, $k = 1$, $K = 2$ and $c_1 = 0.1$.

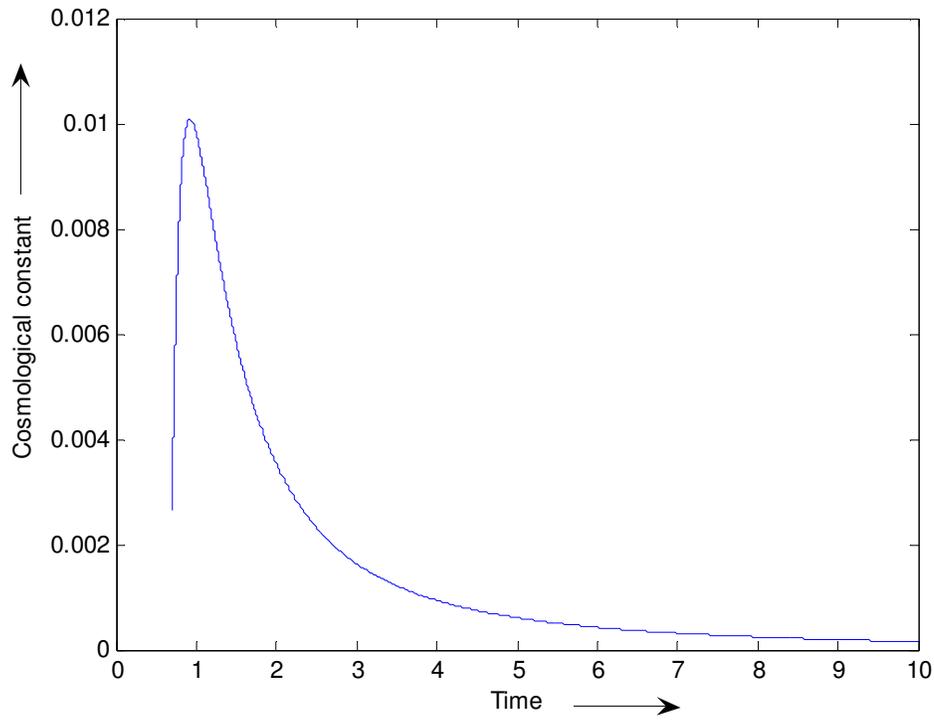

Fig.2. The plot of cosmological constant $\Lambda$ versus time t for the model (42) with parameters $\gamma = 0.5$, $\alpha = 0.2$, $n = 1$, $k = 1$, $K = 2$ and $c_1 = 0.1$.

From equation (47), we observe that at the time of early universe the cosmological constant ($\Lambda$) is negative (t < 0.678) and it increases rapidly during a very short period of time (see Fig.1) For t = 0.678, the value of cosmological constant becomes positive and get its maximum value at t =0.9275 then it decreases as time increases (see Fig. 2). We also observe that the value of $\Lambda$ is small and positive at late time which is supported by recent type Ia supernovae observation[39-41]. The various values of $\Lambda$ at different cosmic time t are given

| S.No. | Cosmic time (t) | Cosmological constant ($\Lambda$) | S.No | Cosmic time (t) | Cosmological constant ($\Lambda$) |
|---|---|---|---|---|---|
| 1 | 0 | -5798 | 9 | 1.183 | 0.008332 |
| 2 | 0.00249 | -5003 | 10 | 1.487 | 0.005929 |
| 3 | 0.00514 | -4292 | 11 | 2.105 | 0.003208 |
| 4 | 0.504 | -0.07598 | 12 | 3.951 | 0.0009643 |
| 5 | 0.678 | 0.0000615 | 13 | 10.00 | 0.0001554 |
| 6 | 0.700 | 0.00264 | 14 | 20.00 | 0.00003923 |
| 7 | 0.817 | 0.00918 | 15 | 50.00 | 0.000006388 |
| 8 | 0.927 | 0.01008 | 16 | 100.00 | 0.000001582 |

The scalar curvature has the following expression

$$R = 6k^2(2-n)(nkt+c_1)^{-2} + \frac{2K^2}{9}(nkt+c_1)^{-\frac{6}{n}} - 6\alpha^2(nkt+c_1)^{-\frac{2}{n}} \qquad (48)$$

The directional Hubble's parameters $H_1$, $H_2$ and $H_3$ are given by

$$H_1 = k(nkt+c_1)^{-1} \qquad (49)$$

$$H_2 = k(nkt+c_1)^{-1} + \frac{K}{3}(nkt+c_1)^{-\frac{3}{n}} \qquad (50)$$

$$H_2 = k(nkt+c_1)^{-1} + \frac{K}{3}(nkt+c_1)^{-\frac{3}{n}} \qquad (51)$$

Where as the average generalized Hubble's parameter is given by

$$H = k(nkt+c_1)^{-1} \qquad (52)$$

The other physical quantities $\theta$, $A_m$ and $\sigma^2$ is given by

$$\theta = 3k(nkt+c_1)^{-1}, \qquad (53)$$

$$A_m = \frac{2}{27}\frac{K^2}{k^2}(nkt+c_1)^{\frac{2n-6}{n}}, \qquad (54)$$

$$\sigma^2 = \frac{K^2}{9}(nkt+c_1)^{-\frac{6}{n}}, \qquad (55)$$

### 4.2 When n=0

In this case, the solutions for A(t), B(t) and C(t) from equation (36)-(38) with help of equation (24) can be given as

$$A(t) = c_2 \exp(kt), \tag{56}$$

$$B(t) = Mc_2 \exp\left[kt - \frac{K}{3c_2^3 k}\exp(-kt)\right], \tag{57}$$

$$B(t) = M^{-1}c_2 \exp\left[kt + \frac{K}{3c_2^3 k}\exp(-kt)\right], \tag{58}$$

Hence the metric (3) reduces to new form

$$ds^2 = dt^2 - c_2^2 \exp(2kt)dx^2 - \exp(2\alpha x)\left[\begin{array}{l} M^2 c_2^2 \exp(2kt - \frac{2K}{3c_2^3 k}\exp(-kt)dy^2 + \\ M^{-2}c_2^2 \exp(2kt + \frac{2K}{3c_2^3 k}\exp(-kt)dz^2 \end{array}\right],$$

(59)

For this derived model (59), the pressure, energy density and cosmological constant are given by

$$8\pi(p + \Lambda) = -3k^2 - \frac{K^2}{9c_2^6}\exp(-2kt) + \frac{m^2}{c_2^2}\exp(-2kt) \tag{60}$$

$$8\pi(\rho - \Lambda) = 3k^2 - \frac{K^2}{9c_2^6}\exp(-2kt) - \frac{3m^2}{c_2^2}\exp(-2kt) \tag{61}$$

From equation (45), (60) and (61), we get

$$8\pi(1+\gamma)\rho = -\frac{2K^2}{9c_2^6}\exp(-2kt) - \frac{2\alpha^2}{c_2^2}\exp(-2kt) \tag{62}$$

Eliminating ρ(t) between (61) and (62), we get

$$8\pi(1+\gamma)\Lambda = \frac{\alpha^2(3\gamma+1)}{c_2^2}\exp(-2kt) + \frac{K^2(\gamma-1)}{9c_2^6}\exp(-2kt) - 3(1+\gamma)k^2 \tag{63}$$

From equation (63), we observe that the value of Cosmological constant decreases as the time increases. Thus it is decreasing function of time and approaches small value .Which is supported by the results from resent type Ia supernovae observations Figure 3 clearly shows the behavior of Cosmological constant Λ as the decreasing function of time.

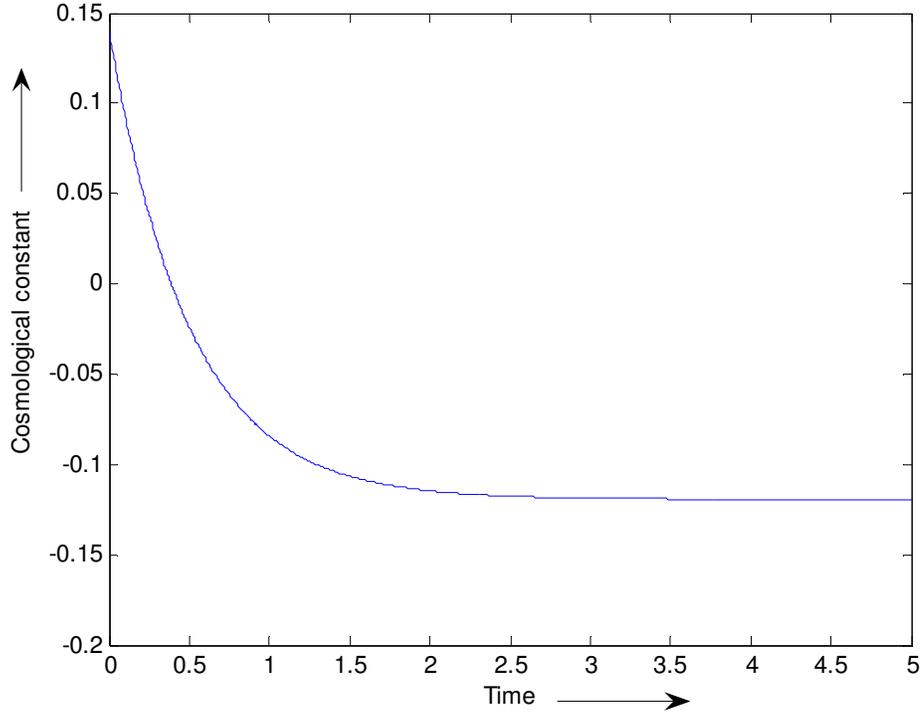

Fig.3. The plot of cosmological constant $\Lambda$ versus time t for the model (59) with parameters $\gamma = 0.5$, $\alpha = 2$, $k = 1$, $K = 2$ and $c_2 = 1$.

From equation (63), we observe that at the time of early universe the cosmological constant ($\Lambda$) is positive (t < 0.39) and it decreases as time increases. For t = 0.39, the value of cosmological constant becomes negative. Thus we observe that the value of $\Lambda$ is small and positive at early time which is supported by recent type Ia supernovae observation[39-41]. As time increases the value of cosmological constant becomes negative which leads the idea of re collapse of our universe. The various values of $\Lambda$ at different cosmic time t are given

| S. No. | Cosmic time (t) | Cosmological Constant ($\Lambda$) | S. No. | Cosmic time (t) | Cosmological Constant ($\Lambda$) |
|---|---|---|---|---|---|
| 1. | 0 | 0.1401 | 8. | 0.35 | 0.0094 |
| 2. | 0.05 | 0.1154 | 9. | 0.385 | 0.00075 |
| 3. | 0.10 | 0.09306 | 10. | 0.39 | -0.00044 |
| 4. | 0.15 | 0.07284 | 11. | 0.40 | -0.002799 |
| 5. | 0.20 | 0.05455 | 12. | 0.45 | -0.0139 |
| 6. | 0.25 | 0.03799 | 13. | 2.50 | -0.1177 |
| 7. | 0.30 | 0.02302 | 14. | 5.00 | -0.1194 |

The scalar curvature has the following expression

$$R = 12k^2 + \left(\frac{2K^2}{9c_2^6} - \frac{6\alpha^2}{c_2^2}\right)\exp(-2kt) \qquad (63)$$

The directional Hubble's parameters $H_1$, $H_2$ and $H_3$ are given by

$$H_1 = k \tag{64}$$

$$H_2 = k + \frac{K}{3c_2^3}\exp(-kt) \tag{65}$$

$$H_2 = k - \frac{K}{3c_2^3}\exp(-kt) \tag{66}$$

Where as the generalized Hubble's parameter is given by

$$H = k \tag{67}$$

The other physical quantities $\theta$, $A_m$ and $\sigma^2$ is given by

$$\theta = 3k \tag{68}$$

$$A_m = \frac{2}{3}\frac{K^2}{k^2 c_2^6}\exp(-2kt), \tag{69}$$

$$\sigma^2 = \frac{K^2}{c_2^6}\exp(-2kt) \tag{70}$$

It is found that the directional Hubble parameters are time dependent while the average Hubble parameter is constant. In this case also we found deceleration parameter q = -1. This is the case of de sitter universe.

## 5. Discussion and Concluding remarks

In this paper we have presented the law of variation for Hubble's parameter in homogeneous and Bianchi type V space-time model that yield a constant value of deceleration parameter. We have obtained exact solutions of EFES for Bianchi type V space-time with a perfect fluid as the source of matter and cosmological term Λ varying with time. It is also seen that solutions obtained by Singh, Shri Ram and Zeyauddin [42] are particular of our solutions. The cosmological constant for model (42) is decreasing function of time and it approach a small positive value at late time. Which are supported by the results from recent supernovae observations obtained by the High-z Supernova Team and Supernova Cosmological project (Garnavich et al [38], Perlmutter et al [39], Rices et al [40], Schmidt et al [41]). The cosmological constant for model (59) is decreasing function of time and it approach a small positive value at early time of universe After that its value is negative (fig 3). In concequence today the estimation of Λ is not only complicated but also uncertain and indirect. It appears however the Einstein- Maxwell theory points to a different approach possibly simpler and more direct, since it consist in measuring a constant electromagnetic background of universe. Such a possibility is illustrated below for Λ≤ 0 [43, 44] ie for the case when the presence of Λ decelerates the expansion of universe. A negative cosmological constant adds to the attractive gravity, therefore universe with negative cosmological constant are invariably doomed to re collapse. For model (59) we observe that initially the value of cosmological constant is positive and its value becomes negative as time increases which signifies the re collapse of universe and generates the next one.

It is possible to discuss entropy of our universe. In thermodynamics the expression for entropy is given by

$$Tds = d(\rho V) + PdV \tag{71}$$

Where V is the spatial volume.

To solve the entropy problem of the standard model, it is necessary to treat ds > 0, for at least the part of evolution of universe. Hence from equation (71)

$$Tds = \rho_4 + (\rho + P)\left(\frac{A_4}{A} + \frac{B_4}{B} + \frac{C_4}{C}\right) > 0 \tag{72}$$

The conservation equation $T_{ij}^{j} = 0$; for metric (3) lead to

$$\rho_4 + (\rho + P)\left(\frac{A_4}{A} + \frac{B_4}{B} + \frac{C_4}{C}\right) + \frac{3}{2}\Lambda\Lambda_4 + \frac{3}{2}\Lambda^2\left(\frac{A_4}{A} + \frac{B_4}{B} + \frac{C_4}{C}\right) = 0 \tag{73}$$

Equation (72) and (73) leads to

$$\Lambda_4 + \Lambda\left(\frac{A_4}{A} + \frac{B_4}{B} + \frac{C_4}{C}\right) < 0 \tag{74}$$

From equation (14) and (74), we have

$$\Lambda_4 + 3\Lambda H < 0 \tag{75}$$

For $n \neq 0$ and n = 0 equation (75) leads to

$$\Lambda < \frac{n_0}{(nkt + c_1)^{\frac{3}{n}}}, \qquad n \neq 0 \tag{76}$$

$$\Lambda < \exp(k_0 - 3kt), \qquad n = 0 \tag{77}$$

Where $n_0$ and $k_0$ are the constant of integration. It is clear that $\Lambda$ is the decreasing function of time in both case. Thus the cosmological constant affect entropy because for entropy ds > 0 leads to the observational result of $\Lambda(t)$. From equation (75) we conclude that cosmological constant is responsible for expansion of universe. For model (42) we find singularity at $t = -\frac{c_1}{3k}$. The rate of expansion, the mean anisotropy parameter and shear scalar all diverse at $t = -\frac{c_1}{3k}$ for n < 3. Also $\lim_{t \to \infty} \frac{\sigma^2}{\theta^2} = 0$, hence model approaches isotropic at late time whereas the model (59) represent singularity free model. For this model also we observe that $\lim_{t \to \infty} \frac{\sigma^2}{\theta^2} = 0$, hence model approaches isotropic at late time. The rate of expansion of universe is uniform through out the evolution. As $t \to \infty$, p = -ρ, which may be considered as vacuum energy density. This class of solution is consistent with the recent observations of supernovae Ia [39-41] and chaplygin gas models by Kamenshchik et al [45] that require the present universe is accelerating. Generally the models presented in this paper are expanding, shearing and accelerating which become isotropic at later time of evolution. The solution obtained in this paper could give an appropriate description of the evolution of universe. More realistic models may be analyzed using this technique, which may lead to interesting and different physical behaviors of the evolution of universe.